\documentstyle[eqsecnum,aps]{revtex}
\tightenlines
 
\begin{document}
\draft
 
\begin{title}
{Scaling and chaos in periodic approximations to the two-dimensional Ising spin glass}
\end{title}
\author{David A. Huse}
\address{Physics Department, Princeton University, Princeton, NJ 08544}
\author{Lee-Fen Ko}
\address{SpecTran Corporation, 50 Hall Road, Sturbridge, MA 01566}
\date{\today}
\maketitle
 
\begin{abstract}
We approximate a two-dimensional spin glass by tiling an infinite lattice with
large identical unit cells.  The interactions within the unit cell are
chosen at random, just as when one studies finite-size systems with
periodic boundary conditions.  But here the unit cells are instead connected
to form an infinite lattice, so one may examine correlations on all length scales,
and the system can have true phase transitions.  For such approximations
to the Ising spin glass on the 
square lattice, we apply the free-fermion method of Onsager and 
the anticommuting operator approach of Kaufman to obtain 
numerically exact results for each
realization of the quenched disorder.   Each such sample shows one or more
critical points, with the distribution of critical temperatures scaling
with the unit cell size, consistent with what is expected from the
scaling theory of low-dimensional spin glasses.  
Due to ``chaos'', the correlations between unit cells can
change sign with changing temperature.  We examine the scaling
of this chaos with unit cell size.  Many samples have multiple 
critical points due to the interactions between unit cells changing sign
at temperatures within the ordered phases.
\end{abstract}
 
\section{Introduction}
 
The numerical investigation of the equilibrium, low-temperature properties of 
finite-dimensional spin-glasses has relied heavily on studies of models on
finite-sized lattices, usually with periodic boundary 
conditions.\cite{BM1,McM,BY,KS,KHS,SL,REA,NY}
The data for finite lattices are extrapolated to the
infinite lattice limit, typically using scaling assumptions.  In this
paper we explore a somewhat different approach to approximating
an infinite two-dimensional Ising spin glass.  Instead of putting
periodic boundary conditions on the finite-size spin glass sample,
we instead ``tile'' the infinite lattice with identical unit cells, each
containing the same realization of the random exchanges.  For planar
lattices this produces a two-dimensional Ising model, various properties of
which can be obtained to numerical precision using algorithms whose time 
and memory requirements grow only as a power of the area of the unit cell.
In particular, we examine the square lattice model with only
nearest-neighbor interactions that are independently and identically
distributed according to a continuous probability distribution.

What should we expect for the behavior of this system?  For higher
temperature, when the correlation length is less than the unit cell
size, the periodicity of the lattice is not relevant and the behavior
should be that of a spin glass.  However, when the correlation
length exceeds the unit cell size, the behavior should cross over
to that of a specific periodic Ising model, which will
have a phase transition in two dimensions.  The resulting ordered
phase may be ferromagnetic between unit cells, so the local 
magnetization patterns in each unit cell are identical and aligned, 
or antiferromagnetic, so adjacent unit cells have
opposite local magnetization patterns.  In either case, the
magnetization pattern within a unit cell is that of a spin glass:
a random pattern that is determined in all its details by the temperature
and the specific quenched random interactions in that sample.  
By calculating the spectrum of each specific sample's transfer matrix, we
can locate its phase transitions and determine whether they
are into ferro- or antiferromagnetic phases.   Even in the disordered
phase, we can determine whether the correlations between adjacent unit
cells are more ferro- or antiferromagnetic by comparing the gaps
in the spectrum at the center and edge of the Brillouin zone.  

For each sample, we obtain the spectrum of its transfer matrix, and
thereby the correlation lengths for ferro- and antiferromagnetic
ordering.  The phase transitions are located by finding where the
gap below the highest ``level'' in the spectrum vanishes, corresponding to 
an infinite correlation length.  Some samples show multiple phase transitions,
because the effective interaction between adjacent unit cells
changes sign with changing temperature.  We identify the highest
transition temperature as $T_c$.  In the limit of an infinite
unit cell, we have the two-dimensional spin glass that has $T_c = 0$
and a correlation length that behaves as $\xi \sim T^{-\nu}$ at
low temperature $T$.  But for unit cells of linear size $L$, the
system ``realizes'' that it is periodic when the correlation length exceeds $L$,
so it is reasonable to expect that $T_c \sim L^{-1/\nu}$.  We measure
the distributions of $T_c$ by examining many samples for
each value of $L$ we study, and find that
the distributions do scale with $L$ as expected, with $\nu \cong 2.7$.

The more novel and striking results of our study are those addressing
the so-called ``chaos'' in the spin-glass correlations.  
The effective interaction between two well-separated spins
in a spin-glass is due to many competing (frustrated) chains of interaction
through intermediate spins.  The precise balance between the energetic
and entropic contributions to these interactions changes with changing
temperature.  The net result of this is that the sign of the correlation
between the two distant spins can change as the temperature is 
changed.\cite{FH,BM}  In our specific geometry this can be readily seen
as the sign of the correlations between equivalent spins in different
unit cells of a specific sample changing between ferro- and 
antiferromagnetic as the temperature is changed.  When these sign changes
occur below $T_c$ they cause additional phase transitions, since they
change the nature of the ordered phases between ferro- and
antiferromagnetic.  This is seen in many samples.  We can also detect
the sign changes when they occur above $T_c$ by monitoring which of the
two correlation lengths (ferro- or antiferromagnetic) is longer.
Our new numerical approach to this problem
allows a much more direct study of this temperature-driven chaos
than was possible in the previous Monte-Carlo study by Ney-Nifle and
Young.\cite{NY}  We find that the number of sign changes grows
rapidly with increasing unit cell size, as is expected from scaling
arguments.
 
\section{The model and the method}

We consider an infinite square lattice Ising model tiled by a regular square
pattern of square unit cells of size $L\times L$ spins, $L$ even. 
We take the pattern of quenched random interactions 
in the upper $L \times (L/2)$ half of the
unit cell to be the mirror image of that in the lower half produced by
reflecting about a horizontal line passing through the central
row of spins; this makes the
the transfer matrix that takes one vertically across a row
of full unit cells Hermitian and hence easier to handle numerically.
The reduced bond strength of the nearest-neighbor bond between
sites $i$ and $j$ is $K_{ij} = J_{ij}/T$, where $J_{ij}$ is a random number
drawn from a distribution that is uniform on the two intervals
(-1.6,-0.4) and (0.4,1.6), and $T$ is the temperature.   
In general, this bond is duplicated in the other half of
the unit cell and in all other unit cells throughout the infinite
lattice.  The unusual bond strength distribution we use was chosen
because the algorithm has difficulty with both very strong and very
weak bonds.  However, since the distribution is broad and continuous,
we expect behavior similar to that for more standard continuous distributions,
such as the Gaussian, that include weak bonds.

The partition function $Z=V^M$, $M\to\infty$, where $V=WW^\dagger$ is 
the transfer matrix across a row of full unit cells and $W$ is that across
a row of half unit cells. The transfer matrix $W$ is a product of row-to-row
transfer matrices of horizontal bonds, $V_h$'s, and vertical bonds, $V_v$'s.
The $V_h$'s and $V_v$'s have the usual quadratic form in terms of the 
anticommuting operators,\cite{ON,KA,KF,AKS} except that the couplings vary from
site to site, both within and between rows. Note that as 
$K_{ij}\to -K_{ij}$,  $ K_{ij}^*\to K_{ij}^*+i\pi$.
The quadratic form allows the transfer matrices
to be represented as matrices, $R$'s, in the space of anticommuting operators, which
is of dimension twice the width of the lattice. This transforms
a problem exponential in the size of the lattice to one that is
linear. In other words, the problem of finding
the eigenvalues, $\Lambda_i$'s, of the transfer matrix is transformed to 
that of computing the
energy levels of the free fermions.  In this free-fermion representation, the 
eigenvalues of the transfer matrix come in pairs of 
$\exp \pm \lambda_i(\omega)$, $i=1,...,L$, 
for each spatial frequency 
$\omega\in(-\pi,\pi)$; the energy levels of the free
fermions are simply $\pm\frac12\lambda_i$'s. In the ground state, all negative
energy levels are filled. The first excited state consists of an excitation at
the gap where $\lambda_1$ is minimum.   Thus the correlation length $\xi$ 
along the vertical direction in
units of the unit cell height is $[\min_{\omega} \lambda_1(\omega)]^{-1}$. 
The special structure of the eigenvalues is the result of each of the $R$'s 
being a direct product of two-dimensional rotations.

The algorithm consists of multiplying the $2L\times 2L$ matrices $R(\omega)$'s from 
each row of horizontal and vertical bonds through $L/2$ rows, then multiplying
by its hermitian conjugate for the mirrored half of the $L/2$ rows. The gap is 
then computed from the smallest eigenvalue above one. Because the entries in
these matrices are hyperbolic functions, high precision is required for
any reasonable size unit cell; we used an arbitrary precision 
software\cite{CLN} to do the computation.  

More details about calculating the properties of Ising models with
inhomogeneous unit cells will be reported elsewhere.\cite{KF2}
 
\section{phase transitions}

For each sample we have calculated the gaps below the largest
eigenvalue of $\log(V)$ for ``excitations'' with wavenumbers
$\omega = 0$ and $\omega = \pi$, at multiple temperatures.
This gap is equal to $L/\xi$, where $\xi$ is the
``ferromagnetic'' (for $\omega = 0$) or
``antiferromagnetic'' (for $\omega = \pi$) correlation length.
Each sample shows one or more phase transitions where 
one of these correlation lengths diverges, so the corresponding
gap vanishes.  Because the model is a planar Ising model
that is in the free-fermion class, we expect that all the
phase transitions are of the usual two-dimensional Ising model
universality class.  Our results do appear to be consistent
with this expectation, although we have not made thorough
checks.  The model is periodic and in its ordered
phases has correlations that are either ferromagnetic or
antiferromagnetic between equivalent spins in adjacent unit 
cells along the rows.  Within a unit cell the ordering is
into a spin-glass magnetization pattern determined by the
random couplings and the temperature.
When a ferromagnetic phase transition occurs, the gap at
$\omega = 0$ vanishes, while that at $\omega = \pi$ vanishes
at an antiferromagnetic transition.  Note, we define the
wavenumber $\omega$ in terms of the unit cell size, so
$\omega = \pi$ corresponds to a period
of two unit cells along the rows.  For a number of samples
we have also examined the gaps at other wavenumbers; these
gaps are never found to vanish.  Because we have chosen the unit
cells to be vertically reflection-invariant, the correlations between
equivalent spins in adjacent unit cells in 
different (horizontal) rows are always ferromagnetic
(the eigenvalues of the transfer matrix $V$ are all real and 
positive).

An example of the most common type of sample, with just one
phase transition, is shown in Fig. 1.  The phase transition is
clearly seen as the point where the gap at $\omega = \pi$
vanishes.  For this sample
the ground state is antiferromagnetic between unit cells
(although samples with ferromagnetic ground states are equally common), 
and the interaction between the unit cells remains
antiferromagnetic for temperatures below the phase transition.  However, 
the net interaction between unit cells may
change sign with changing temperature as the balance between
energetic and entropic interactions is changed in the full
free energy.  For the sample of Fig. 1 this happens in the disordered
high temperature phase, near $1/T = 1.2$. 

In some samples (between 10 and 20\% of the samples for $20 \leq L \leq 40$)
the sign of the intercell interaction changes at low enough
temperature to produce multiple phase transitions.  An example
is shown in Fig. 2.  Here the ground state is again antiferromagnetic,
but the intercell interaction changes sign at low
temperature (near $1/T = 3$).  This low temperature transition
region is shown with expanded scales in Fig. 3.  There is a
narrow interval of disordered phase where the interaction
between horizontally-adjacent unit cells in a row is very small or zero.  
Note that the minimum gap remains less than about 0.1 in this
low-temperature disordered phase.  This means the vertical
correlation length always exceeds about 10 unit cells,
which is 300 nearest-neighbor spacings in this case.  Thus
this is a very well correlated disordered phase, as is to
be expected at such a low temperature.
This disordered phase is bounded below by the low-temperature
antiferromagnetic phase and above by an intermediate-temperature
ferromagnetic phase.  The ferromagnetic phase disorders in the
third phase transition at higher temperature (Fig. 2).  We have seen
samples that show as many as 5 transitions in the temperature range 
we studied, but these are rather rare (roughly 40 in a 
total population of about 14,000 samples examined).

Our system should behave like a spin glass as long as the correlations
between unit cells are weak.  However, once the correlation length
reaches the unit cell size $L$ the system ``realizes'' that it is
periodic and can order as a regular two-dimensional Ising model.
The temperature, $T_c$, of the first phase transition met
on lowering the temperature thus indicates where the
correlation length is of order $L$.  The correlation length in the
two-dimensional spin glass diverges as $T^{-\nu}$ at 
low temperature.\cite{BM1,McM}
Thus we expect the mean or median transition temperature for our
periodic approximations to the spin glass to decrease with
increasing $L$ as $L^{-1/\nu}$.  For a small minority of samples 
the transition is at 
a lower temperature than we studied, so we have only an upper bound on it.
Because of this we do not have a proper measurement of the mean $T_c$,
but we do have the medians for each size, $T_{med}(L)$.  
The results are $T_{med}(20) = 0.757 \pm 0.003$ from 5000
samples, $T_{med}(30) = 0.654 \pm 0.003$ from 5000 samples, and
$T_{med}(40) = 0.588 \pm 0.003$ from 4000 samples.  These give
an estimate of $\nu \cong 2.7$ from fitting to
$T_{med}(L) \sim L^{-1/\nu}$.  The distribution of
$T_c/T_{med}(L)$ should be independent of $L$, by scaling.
This is demonstrated in Fig. 4, where the histograms of
the scaled $T_c$ are shown.

Previous estimates of $\nu$ for the two-dimensional Ising spin
glass with a continuous distribution of bond strengths have ranged 
from near 2.0,\cite{KHS,SL,REA,NY}
to close to 4, \cite{REA} with the estimates depending
on the scaling assumptions used to obtain them.    
Our estimate of 2.7 is comfortably within this range.
Note that we have used data for only a factor of two change
in $L$ for this estimate, so it could be substantially
affected by corrections to scaling that we would not detect.
For smaller $L$ (e.g., $L=10$) we saw clear deviations from
scaling, for example in the shape of the mean gap 
vs. $T$ (Fig. 5), so we have
not used the smaller $L$ data.  The unusual distribution of
interactions we use is a sort of hybrid of the $\pm$J and
continuous distributions, so could show some sort of
crossover at short length scales.

Different researchers using the same assumptions have generally
good agreement on the value of $\nu$ for Gaussian-distributed
interactions, so the differences do not appear to arise from
variable quality in the underlying numerical
data.  The wide variation of estimates of $\nu$ indicates
that there is something about the low-temperature scaling
in the two-dimensional Ising spin glass that is not yet understood.

We have also averaged the minimum of the gaps (at $\omega = 0$ or $\pi$)
over all samples for each $L$ and various temperatures.  
The gap is $L/\xi$, so is dimensionless, and
thus its average should be independent of $L$ at a fixed
$T/T_{med}(L)$.  The resulting scaling plot of the average gap is
shown in Fig. 5.  The scaling collapse is quite good near and
below $T_{med}$, but clear deviations from scaling can be
seen at higher temperatures.

\section{``chaos''}

The term ``chaos'' is used to refer to the extreme sensitivity of
the sign of the long-distance spin-spin correlations to changes
of temperature in a given spin glass sample.\cite{BM}  One manifestation
of these sign changes that we have already discussed is the multiple
phase transitions that occur in some samples.  To make a more
quantitative study of this we define the coupling between adjacent
unit cells to be ferro- or antiferromagnetic based on whether the
gap is smaller at $\omega = 0$ or $\omega = \pi$, respectively.
Then we record each time this coupling changes sign as the temperature
is changed for each sample.  Let $n_L(T)$ be the number of sign
changes per sample between zero temperature and temperature $T$ for
unit cell size $L$.  We do not directly measure $n_L(T)$, since we
do not follow the samples all the way to $T=0$, but we do measure
the change in $n_L(T)$ over the temperature range we study. 
In Fig. 6 we show $n_L(T) - n_L(T_{min})$ vs. $T$, where $T_{min}$ 
is the lowest temperature we examined for each size $L$.  It is clear 
from this figure that the number of coupling
changes increases with increasing $L$.  

Why does the coupling change sign with changing $T$?  This occurs
because for nonzero $T$ and large enough $L$ the free energy 
difference between ferro- and antiferromagnetic
spin patterns is a near-cancellation of larger entropy and energy
differences.  The free energy difference varies with $L$ as 
$\Delta F \sim L^{-1/\nu}$,
so actually decreases with increasing $L$ in this two-dimensional system.
At low temperatures, $T << T_c$,
the difference between the two patterns is typically a relative domain
wall running across the full unit cell.  The two different spin configurations
along that domain wall will have randomly different local entropies 
due to fluctuations at small length scales.  
Thus it has been argued \cite{FH,BM} that the
entropy difference between the two patterns grows 
with $L$ as $\Delta S \sim L^{d_s/2}$,
where $d_s$ is the fractal dimension of the relative domain wall.
For this case $1 \leq d_s \leq 2$, the lower limit being that of
a straight wall, and the upper limit being a space-filling rough wall.
Since the free energy difference is $\Delta F = \Delta E - T\Delta S$,
where $\Delta E$ is the energy difference, the typical temperature
change needed to change the sign of $\Delta F$ varies with L as
$\delta T \sim \Delta F/\Delta S \sim L^{-d_s/2 - 1/\nu}$.  Thus the
rate of sign changes with changing $T$ is expected to increase strongly
with increasing $L$, as we show in Fig. 6.  We believe this is the
first direct measurement of these sign changes of the correlations
in a realistic spin glass model at low temperatures.

The above argument applies at low temperatures where the system is
well-correlated within a unit cell, so it is reasonable to assume
the difference between the two spin patterns is simply a 
relative domain wall.  This picture will break down at higher
temperatures as $T$ approaches and exceeds $T_c$.  
To take account of this, a naive finite-size scaling hypothesis is

\begin{equation}
n_L(T) \sim L^{\zeta} T_{med}(L) N(\frac{T}{T_{med}(L)}) ,
\end{equation}
 
\noindent where $N$ is a scaling function, and the chaos exponent $\zeta$
determines how the scaled chaos increases with
increasing $L$.  In Fig. 7, we show the data scaled in this
way, with $\zeta = 0.85$.  As with our estimate of $\nu$, this
estimate of $\zeta$ is from scaling over a factor of only two in
length.  There are presumably errors in these exponent estimates 
due to corrections to scaling, which are likely to be much larger
than those due to the statistical uncertainties in our data.  
Thus we do not know how large an error estimate to quote
for our exponent estimates.
Our estimate of $\zeta \cong 0.85$ is consistent with the
$\zeta = 1.0 \pm 0.2$ found by Ney-Nifle and Young\cite{NY} from 
overlaps between different temperatures fit to a similar scaling form.  
In the latter Monte-Carlo study the overlap was only suppressed by a 
few per cent by the temperature change, indicating that only a 
very small fraction of the
spins were flipped by the temperature changes in the rather small
($L \leq 10$) samples they could study using that approach.  In
contrast, we are able to access substantially
larger length scales and a wider temperature
range, and find at least one sign change in the correlations for
a large fraction of the samples.

\begin{center}
{FIGURE CAPTIONS}
\end{center}

\noindent {\bf FIG. 1.}   The gaps below the highest state in the spectrum
of the logarithm of the transfer matrix for
various wavenumbers vs. the inverse temperature ($1/T$) for a specific  
sample with unit cell size $L = 30$.  At low temperature, this
sample orders antiferromagnetically, as indicated by the gap at $\omega = \pi$
vanishing at the phase transition.  However, the minimum gap moves to
$\omega = 0$ at the highest temperature point shown.

\noindent {\bf FIG. 2.}   As in Fig. 1, the gap vs. $1/T$, but 
here for a different specific sample with $L=30$ that 
instead shows three phase transitions.  Moving from
left to right (high to low temperature) the phases are: disordered,
ferromagnetic, disordered again, and antiferromagnetic.  The intercell
interaction changes sign near $1/T = 3$ from ferromagnetic at higher
temperature to antiferromagnetic at lower temperature.

\noindent {\bf FIG. 3.}  The lower-temperature transition region in Fig. 2,
shown with expanded scales.

\noindent {\bf FIG. 4.}  The histograms of the scaled transition temperatures,
$T_c/T_{med}(L)$ for each size $L$, where $T_{med}(L)$ is the median
transition temperature for unit cell size $L$.  The bins in these histograms are 
all of width 0.1 along the horizontal axis, and the fraction of the
samples lying within each bin is indicated.

\noindent {\bf FIG. 5.}  The mean of the gap below the highest
state in the spectrum of the
logarithm of the transfer matrix vs. the temperature scaled by the
median transition temperature, $T_{med}(L)$, for each size $L$.

\noindent {\bf FIG. 6.}  The number of sign changes in the correlations
between adjacent unit cells per sample between the lowest temperature
studied, $T_{min}$, and temperature $T$.  For sizes $L = 20, 30, 40$
we used $1/T_{min} = 3.4, 3.8, 4.2$, respectively.

\noindent {\bf FIG. 7.}  The number of sign changes per sample, scaled
according to Eq. (4.1), with $\zeta = 0.85$.
 

\begin{references}
 
\bibitem{BM1} A. J. Bray and M. A. Moore, J. Phys. C {\bf 17}, L463 (1984).

\bibitem{McM} W. L. McMillan, Phys. Rev. B {\bf 30}, 476 (1984).

\bibitem{BY}  R. N. Bhatt and A. P. Young, Phys. Rev. B {\bf 37},
5606 (1988).
 
\bibitem{KS} N. Kawashima and M. Suzuki, J. Phys. A {\bf 25},
1055 (1992).
 
\bibitem{KHS} N. Kawashima, N. Hatano and M. Suzuki, J. Phys
A {\bf 25}, 4985 (1992).
 
\bibitem{SL} S. Liang, Phys. Rev. Lett. {\bf 69}, 2145 (1992).
 
\bibitem{REA} H. Rieger, {\it et al.}, J. Phys. A {\bf 29}, 3939 (1996).
 
\bibitem{NY} M. Ney-Nifle and A. P. Young, J. Phys. A [in press].
 
\bibitem{FH} D. S. Fisher and D. A. Huse, Phys. Rev. Lett. {\bf 56}, 1601
(1986); Phys. Rev. B {\bf 37}, 373 and 386 (1988).

\bibitem{BM} A. J. Bray and M. A. Moore, Phys. Rev. Lett. {\bf 58}, 57 (1987).

\bibitem{ON} L. Onsager, Phys. Rev. {\bf 65}, 117 (1944).

\bibitem{KA} B. Kaufman, Phys. Rev. {\bf 76}, 1232 (1949).

\bibitem{KF} L.-F. Ko and M. E. Fisher, J. Stat. Phys. {\bf 58}, 249 (1990).

\bibitem{AKS} D. B. Abraham, L.-F. Ko and N. M. Svraki\'c, J. Stat. Phys.
{\bf 56}, 563 (1989).

\bibitem{CLN} B. Haible, ``CLN, a class library for numbers'', available
from ftp://ma2s2.mathematik.uni-karlsruhe.de/pub/gnu/cln/tar.z (1996).

\bibitem{KF2} L.-F. Ko and M. E. Fisher, (to be published).
 
\end{references}
\end{document}